\PassOptionsToPackage{unicode}{hyperref}
\PassOptionsToPackage{hyphens}{url}
\PassOptionsToPackage{dvipsnames,svgnames,x11names}{xcolor}
\documentclass[
  letterpaper,
  DIV=11,
  numbers=noendperiod]{scrartcl}

\usepackage{amsmath,amssymb}
\usepackage{lmodern}
\usepackage{iftex}
\ifPDFTeX
  \usepackage[T1]{fontenc}
  \usepackage[utf8]{inputenc}
  \usepackage{textcomp} 
\else 
  \usepackage{unicode-math}
  \defaultfontfeatures{Scale=MatchLowercase}
  \defaultfontfeatures[\rmfamily]{Ligatures=TeX,Scale=1}
\fi
\IfFileExists{upquote.sty}{\usepackage{upquote}}{}
\IfFileExists{microtype.sty}{
  \usepackage[]{microtype}
  \UseMicrotypeSet[protrusion]{basicmath} 
}{}
\makeatletter
\@ifundefined{KOMAClassName}{
  \IfFileExists{parskip.sty}{%
    \usepackage{parskip}
  }{
    \setlength{\parindent}{0pt}
    \setlength{\parskip}{6pt plus 2pt minus 1pt}}
}{
  \KOMAoptions{parskip=half}}
\makeatother
\usepackage{xcolor}
\ifx\paragraph\undefined\else
  \let\oldparagraph\paragraph
  \renewcommand{\paragraph}[1]{\oldparagraph{#1}\mbox{}}
\fi
\ifx\subparagraph\undefined\else
  \let\oldsubparagraph\subparagraph
  \renewcommand{\subparagraph}[1]{\oldsubparagraph{#1}\mbox{}}
\fi

\usepackage{longtable,booktabs,array}
\usepackage{calc} 
\usepackage{etoolbox}
\makeatletter
\patchcmd\longtable{\par}{\if@noskipsec\mbox{}\fi\par}{}{}
\makeatother
\IfFileExists{footnotehyper.sty}{\usepackage{footnotehyper}}{\usepackage{footnote}}
\makesavenoteenv{longtable}
\usepackage{graphicx}
\makeatletter
\def\maxwidth{\ifdim\Gin@nat@width>\linewidth\linewidth\else\Gin@nat@width\fi}
\def\maxheight{\ifdim\Gin@nat@height>\textheight\textheight\else\Gin@nat@height\fi}
\makeatother
\setkeys{Gin}{width=\maxwidth,height=\maxheight,keepaspectratio}
\makeatletter
\def\fps@figure{htbp}
\makeatother
\newlength{\cslhangindent}
\newlength{\csllabelwidth}
\newlength{\cslentryspacingunit} 
\newenvironment{CSLReferences}[2] 
 {
  \setlength{\parindent}{0pt}
  \ifodd #1
  \let\oldpar\par
  \def\par{\hangindent=\cslhangindent\oldpar}
  \fi
  \setlength{\parskip}{#2\cslentryspacingunit}
 }%
 {}
\usepackage{calc}

\newcommand{\CSLLeftMargin}[1]{\parbox[t]{\csllabelwidth}{#1}}
\newcommand{\CSLRightInline}[1]{\parbox[t]{\linewidth - \csllabelwidth}{#1}\break}

\usepackage{booktabs}
\usepackage{longtable}
\usepackage{array}
\usepackage{multirow}
\usepackage{wrapfig}
\usepackage{float}
\usepackage{colortbl}
\usepackage{pdflscape}
\usepackage{tabu}
\usepackage{threeparttable}
\usepackage{threeparttablex}
\usepackage[normalem]{ulem}
\usepackage{makecell}
\usepackage{xcolor}
\KOMAoption{captions}{tableheading}
\makeatletter
\@ifpackageloaded{caption}{}{\usepackage{caption}}
\AtBeginDocument{%
\ifdefined\contentsname
  \renewcommand*\contentsname{Table of contents}
\else
  \newcommand\contentsname{Table of contents}
\fi
\ifdefined\listfigurename
  \renewcommand*\listfigurename{List of Figures}
\else
  \newcommand\listfigurename{List of Figures}
\fi
\ifdefined\listtablename
  \renewcommand*\listtablename{List of Tables}
\else
  \newcommand\listtablename{List of Tables}
\fi
\ifdefined\figurename
  \renewcommand*\figurename{Figure}
\else
  \newcommand\figurename{Figure}
\fi
\ifdefined\tablename
  \renewcommand*\tablename{Table}
\else
  \newcommand\tablename{Table}
\fi
}
\@ifpackageloaded{float}{}{\usepackage{float}}
\floatstyle{ruled}
\@ifundefined{c@chapter}{\newfloat{codelisting}{h}{lop}}{\newfloat{codelisting}{h}{lop}[chapter]}
\floatname{codelisting}{Listing}

\makeatother
\makeatletter
\@ifpackageloaded{caption}{}{\usepackage{caption}}
\@ifpackageloaded{subcaption}{}{\usepackage{subcaption}}
\makeatother
\makeatletter
\@ifpackageloaded{tcolorbox}{}{\usepackage[many]{tcolorbox}}
\makeatother
\makeatletter
\@ifundefined{shadecolor}{\definecolor{shadecolor}{rgb}{.97, .97, .97}}
\makeatother
\ifLuaTeX
  \usepackage{selnolig}  
\fi
\IfFileExists{bookmark.sty}{\usepackage{bookmark}}{\usepackage{hyperref}}
\IfFileExists{xurl.sty}{\usepackage{xurl}}{} 
\hypersetup{
  pdftitle={Digitization of the Australian Parliamentary Debates, 1998-2022},
  pdfauthor={Lindsay Katz; Rohan Alexander},
  colorlinks=true,
  linkcolor={blue},
  filecolor={Maroon},
  citecolor={Blue},
  urlcolor={Blue},
  pdfcreator={LaTeX via pandoc}}

\title{Digitization of the Australian Parliamentary Debates, 1998-2022}
\author{Lindsay Katz\footnote{University of Toronto,
  lindsay.katz@mail.utoronto.ca} \and Rohan Alexander\footnote{University
  of Toronto, rohan.alexander@utoronto.ca}}
\date{August 10, 2023}

\begin{document}
\maketitle
\begin{abstract}
Public knowledge of what is said in parliament is a tenet of democracy,
and a critical resource for political science research. In Australia,
following the British tradition, the written record of what is said in
parliament is known as Hansard. While the Australian Hansard has always
been publicly available, it has been difficult to use for the purpose of
large-scale macro- and micro-level text analysis because it has only
been available as PDFs or XMLs. Following the lead of the Linked
Parliamentary Data project which achieved this for Canada, we provide a
new, comprehensive, high-quality, rectangular database that captures
proceedings of the Australian parliamentary debates from 1998 to 2022.
The database is publicly available and can be linked to other datasets
such as election results. The creation and accessibility of this
database enables the exploration of new questions and serves as a
valuable resource for both researchers and policymakers.
\end{abstract}
\ifdefined\Shaded\renewenvironment{Shaded}{\begin{tcolorbox}[sharp corners, breakable, enhanced, borderline west={3pt}{0pt}{shadecolor}, boxrule=0pt, interior hidden, frame hidden]}{\end{tcolorbox}}\fi

\hypertarget{sec-intro}{%
\section{Background \& Summary}\label{sec-intro}}

The official written record of parliamentary debates, formally known as
Hansard\textsuperscript{1}, plays a fundamental role in capturing the
history of political proceedings and facilitating the exploration of
valuable research questions. Originating in the British parliament, the
production of Hansard became tradition in many other Commonwealth
countries, such as Canada and Australia\textsuperscript{2}. Given the
content and magnitude of these records, they have significance,
particularly in the context of political science research. In the case
of Canada, the Hansard has been digitized for 1901 to
2019\textsuperscript{3}. Having a digitized version of Hansard enables
researchers to conduct text analysis and statistical modelling.
Following the lead of that project, in this paper we introduce a similar
database for Australia. This is composed of individual datasets for each
sitting day in the House of Representatives from March 1998 to September
2022, containing details on everything said in parliament in a form that
can be readily used by researchers. With the development of tools for
large-scale text analysis, this database will serve as a resource for
understanding political behaviour in Australia over time.

There are a wide variety of potential applications of this database. For
instance, within Australia there is considerable concern that there has
been a decline in the `quality' of public policy debate (however that
might be defined). Our dataset could be used to look at whether it is
really getting worse in particular ways, and if so, why. We might also
be interested in whether particular sub-populations are appropriately
represented in what is talked about in parliament. For instance, there
is often concern that regional areas are overlooked compared with
metropolitan areas. Again, our database could be used to examine whether
this has changed over time. We have developed our database in such a way
that it could be linked with similar databases from other countries
which would enable comparative analysis. For instance, we may be
interested in how the policy focus of a parliament changes given various
global events such as pandemics or wars. An international linkage
provides a comparison case where domestic issues are different while
international ones are common. As an example of enabling this linkage we
have included PartyFacts IDs (https://partyfacts.herokuapp.com) in our
database. This should make it possible to link our database with other
large parliamentary speech collection projects, such as
ParlaMint\textsuperscript{4}, ParlSpeech\textsuperscript{5},
ParlEE\textsuperscript{6}, and MAPLE\textsuperscript{7}.

The Australian House of Representatives, often referred to as `the
House', performs a number of crucial governmental functions, such as
creating new laws and overseeing government
expenditure\textsuperscript{8, ch.~1}. Politicians in the House are
referred to as Members of Parliament (MPs). The House operates under a
parallel chamber setup, meaning there are two debate venues where
proceedings take place: the Chamber, and the Federation Chamber.
Sittings of the House follow a predefined order of business, regulated
by procedural rules called standing orders\textsuperscript{8, ch.~8}. A
typical sitting day in the Chamber has a number of scheduled proceedings
including debates on government business, 90 second member statements,
and Question Time\textsuperscript{8, ch.~8}. The Federation Chamber was
created in 1994 as a subordinate debate venue of the Chamber. This
allows for better time management of House business as its proceedings
occur simultaneously with those of the Chamber\textsuperscript{8,
ch.~21}. Sittings in the Federation Chamber are different to those of
the Chamber in terms of their order of business and scope of discussion.
Business matters discussed in the Federation Chamber are limited largely
to intermediate stages of bill development, and the business of private
Members\textsuperscript{8, ch.~21}. It is the recording and compilation
of these proceedings on which Hansard is based, and it is essentially,
but not entirely, verbatim.

A week or so after each sitting day, a transcript is available for
download from the official Parliament of Australia website in both PDF
and extensible markup language (XML) form. The PDF is the official
release. The PDF imposes formatting designed for humans to read with
ease, whereas XML is designed for consistency and machine legibility.
The nature of XML enables us to more easily use code to manipulate these
records at scale, motivating our choice to develop our database solely
using the XML formatted files. In cases where we were unsure on how to
proceed with processing the XML, we defer first to the PDF, and then to
the video recording of the proceeding, if available.

At present, the Hansard format that is available on the Parliament of
Australia website is not easily accessible for large scale analysis. To
this point, various researchers have had to create their own databases
of usable, complete data based on content from the Australian Parliament
website. For instance, an online, easy to read database of Hansard from
1901 to 1980 using the XML files has been created by Tim Sherratt
(http://historichansard.net/). These data can be navigated by year,
parliament, people, and bills. To make the Australian Parliamentary
Handbook more accessible, an \texttt{R} package which includes data on
all MPs from 1945 to 2019 has been created by Patrick Leslie
(https://github.com/palesl/AustralianHouseOfRepresentatives). Further,
there is the \texttt{AustralianPoliticians} \texttt{R} package, which
contains several datasets related to the political and biographical
information of Australian federal politicians who were active between
1901 and 2021\textsuperscript{9}. And finally, there has been
examination of speech and MP level data between 1990 and 2019 in
Australia\textsuperscript{10}. Like us they scrape the Hansard record
and link it with biographical data. The key difference is that our focus
is on the database itself, while they are focused on using a database
constructed from the same source to answer a particular question about
speaker time. This different focus leads to different emphasis and
approaches.

Many papers exist which use components of Australian Hansard to explore
various research topics. For example, the Hansard has been used to
investigate occurrences of unparliamentary comments by MPs, where the
Speaker tells that MP to withdraw their remark\textsuperscript{11}.
Question Time data from Hansard transcripts during February and March of
2003 has been used to investigate resistance of politicians in answering
questions about Iraq\textsuperscript{12}. Hansard has also been used to
quantify political prominence by investigating strategic mentions of
interest groups by elected officials\textsuperscript{13}. Finally, a
dataset of the Australian Hansard has been constructed and then used to
analyze the effect of elections and changes in Prime Ministers upon
topics mentioned in parliament\textsuperscript{14}. This was created
with the static PDF versions of Hansard, using OCR to digitize these
files into text which is suitable for analysis. This means there are
considerable digitization errors especially in the first half of the
dataset.

While there is evidently a growing body of literature on this topic,
there is still no comprehensive database for Australian Hansard based on
XML that spans from 1901 to the present day. Our work begins to bridge
this gap.

\hypertarget{methods}{%
\section{Methods}\label{methods}}

Our database contains one comma-separated value (CSV) file and one
parquet file for each sitting day of the House of Representatives from
02 March 1998 to 08 September 2022. We developed four scripts to produce
these files. Each script parses Hansard documents from a specific
portion of the 1998 to 2022 time frame.

This section is structured as follows. First, we provide an overview of
our approach to understanding and parsing an individual Hansard XML
document, which informed the scripts used to create our database. This
will be supplemented with an excerpt from a Hansard XML to provide a
visual example of its structure. Next we will explain the specific
differences between the scripts, and outline what structural changes
necessitated their separate development. We then provide details on the
methodological intricacies of three core components of Hansard
proceedings: Question Time, interjections, and stage directions.
Further, we discuss the script we developed to fill in remaining missing
details on the MP speaking, which each file in our database was passed
to after being parsed and cleaned. Finally, we review the supplementary
Hansard debate topics dataset and supplementary divisions dataset we
created to expand the versatility of our database.

\hypertarget{sec-overview}{%
\subsection{Overview}\label{sec-overview}}

The approach to parsing contents of an XML document depends on its tree
structure. As such, to create this database, we started by looking at a
single Hansard XML transcript from 2019. Doing so enabled us to identify
the various components of interest in the document, and how each one can
be parsed according to its corresponding structural form. Parsing was
performed in \texttt{R} using the \texttt{XML} and \texttt{xml2}
packages\textsuperscript{15,16}. Focusing on one transcript also allowed
us to ensure that all key components of the transcript were parsed and
captured in as much detail as possible. The typical form of a Hansard
XML transcript is summarized in the nested list below. This provides an
overview, but does not contain every possible nested element that may be
found in a Hansard XML.

\begin{quote}
\begin{verbatim}
     <hansard>
       1. <session.header>
       2. <chamber.xscript>
          a)  <business.start>
          b)  <debate>
              i.  <debateinfo>
              ii. <debate.text>
              iii. <speech>
              iv. <subdebate.1>
                  (1) <subdebateinfo>
                  (2) <subdebate.text>
                  (3) <speech>
                  (4) <subdebate.2>
                      (a) <subdebateinfo>
                      (b) <subdebate.text>
                      (c) <speech>
      3. <fedchamb.xscript>
      4. <answers.to.questions>
          a)  <question>
          b)  <answer>
\end{verbatim}
\end{quote}

The outer-most node, also known as the parent node, is denoted
\texttt{\textless{}hansard\textgreater{}} and serves as a container for
the entire document. This parent node may have up to four child nodes,
where the first child node contains details on the specific sitting day.
Next, \texttt{\textless{}chamber.xscript\textgreater{}} contains all
proceedings of the Chamber,
\texttt{\textless{}fedchamb.xscript\textgreater{}} contains all
proceedings of the Federation Chamber, and
\texttt{\textless{}answers.to.questions\textgreater{}} contains Question
Time proceedings. The Federation Chamber does not meet on every sitting
day, so this child element is not present in every XML file. The use of
separate child nodes allows for the distinction of proceedings between
the Chamber and Federation Chamber. The structure of the
\texttt{\textless{}chamber.xscript\textgreater{}} and
\texttt{\textless{}fedchamb.xscript\textgreater{}} nodes are generally
the same, where the proceeding begins with
\texttt{\textless{}business.start\textgreater{}} which is followed by a
series of debates. Debate nodes can contain a
\texttt{\textless{}subdebate.1\textgreater{}} child node which has a
\texttt{\textless{}subdebate.2\textgreater{}} child node nested within
it. That said, sometimes \texttt{\textless{}subdebate.2\textgreater{}}
is not nested within \texttt{\textless{}subdebate.1\textgreater{}}. Each
of these three elements (i.e.~\texttt{\textless{}debate\textgreater{}},
\texttt{\textless{}subdebate.1\textgreater{}}, and
\texttt{\textless{}subdebate.2\textgreater{}}) as well as their
respective sub-elements contain important information on the topic of
discussion, who is speaking, and what is being said. The
\texttt{\textless{}speech\textgreater{}} node within each one contains
the bulk of the text associated with that debate or sub-debate. A
typical \texttt{\textless{}speech\textgreater{}} node begins with a
\texttt{\textless{}talk.start\textgreater{}} sub-node, providing
information on the MP whose turn it is to speak and the time of their
first statement. Unsurprisingly, speeches rarely go uninterrupted in
parliamentary debate settings --- they are often composed of a series of
interjections and continuations. These statements are categorized under
different sub-nodes depending on their nature, such as
\texttt{\textless{}interjection\textgreater{}} or
\texttt{\textless{}continuation\textgreater{}}. The final key component
of Hansard is Question Time, in which questions and answers are
classified as unique elements. More detail on the purpose and processing
of Question Time will follow.

\begin{figure}

{\centering \includegraphics{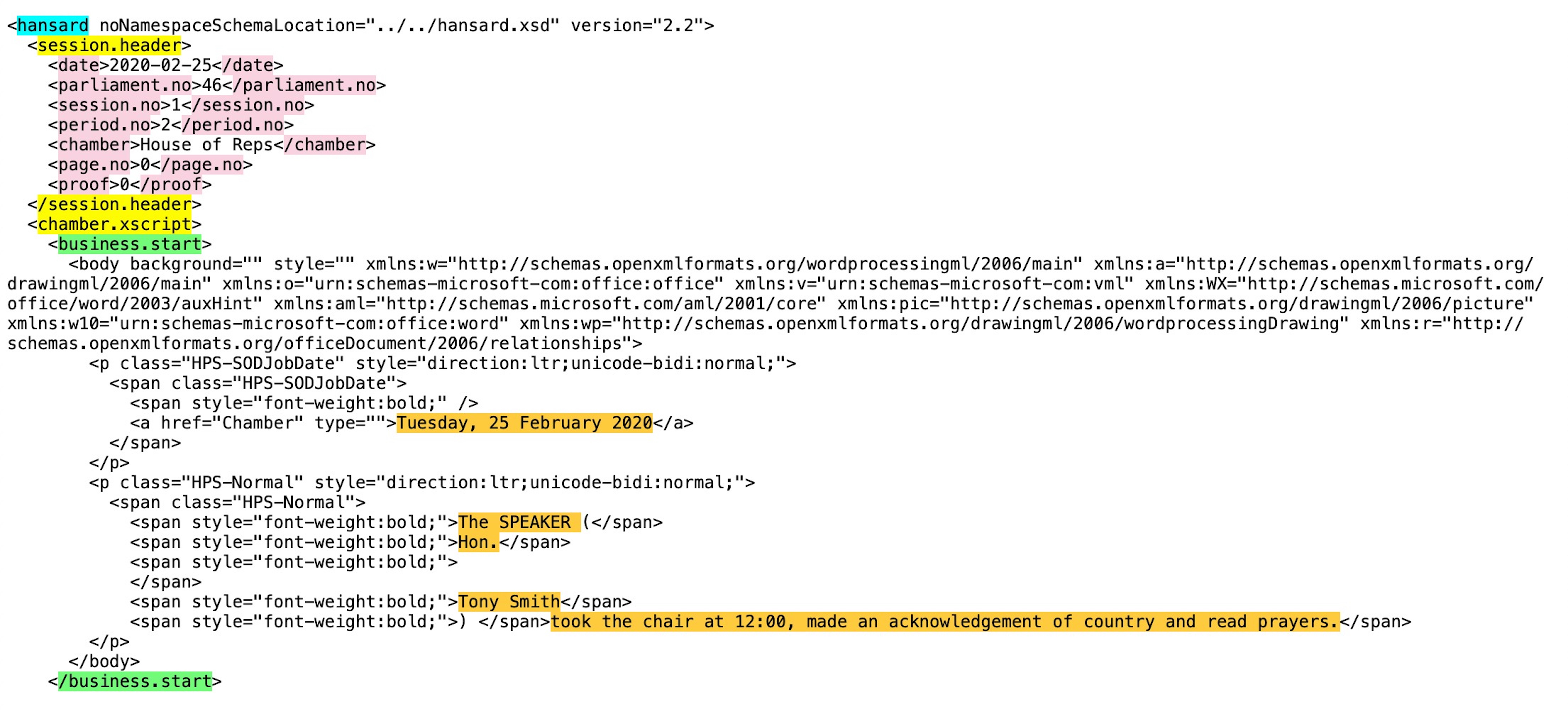}

}

\caption{\label{fig-xml1}Snapshot of the beginning of the XML file for
Hansard on 25 February 2020}

\end{figure}

Figure~\ref{fig-xml1} provides an example of the beginning of an XML
file for Hansard, which illustrates the structure outlined in the nested
list above. As stated, the XML structure begins with a parent element
\texttt{\textless{}hansard\textgreater{}} (highlighted in blue),
followed by a child element
\texttt{\textless{}session.header\textgreater{}} (highlighted in yellow)
with sub-child elements such as the date and parliament number, which
are all highlighted in pink. Next, there is the child element containing
everything that takes place in the Chamber,
\texttt{\textless{}chamber.xscript\textgreater{}}, which is also
highlighted in yellow in Figure~\ref{fig-xml1}. As previously mentioned,
the first sub-node of \texttt{\textless{}chamber.xscript\textgreater{}}
is \texttt{\textless{}business.start\textgreater{}}. The structure of
this can be seen between the nodes highlighted in green in
Figure~\ref{fig-xml1}, where the content we parse from the business
start is highlighted in orange.

Evidently, the nature of XML formatting means that different pieces of
information are categorized under a series of uniquely named and nested
nodes. As a result, to parse each piece of information, one must specify
the unique hierarchy of the nodes in which it is structured. This is
known as an XPath expression, and tells the parser how to navigate the
XML document to obtain the desired information. For example, the session
header date in Figure~\ref{fig-xml1} can be accessed using the XPath
expression ``\texttt{hansard/session.header/date}''. When specifying an
XPath expression, one can use an ``or'' operator to obtain elements from
multiple node paths at once, in the order that they appear in the
document. We did so throughout our script as we parsed uniquely nested
speech content. This allows the correct ordering of elements to be
maintained. We began our first script by parsing all the business start,
speech text, and Question Time contents contained in the XML document,
using these unique XPath expressions to do so.

The next step was to further develop our script to produce tidy data
sets\textsuperscript{17}. These contain all parsed text elements, where
each statement is separated onto its own row with details about the MP
who is speaking, and rows are maintained in chronological order. This
first involved correcting the variable classes and adding several
indicator variables to differentiate where statements came from, such as
Chamber versus Federation Chamber or
\texttt{\textless{}subdebate.1\textgreater{}} versus
\texttt{\textless{}subdebate.2\textgreater{}}. The next key task stemmed
from the fact that the raw text data were not separated by each
statement when parsed. In other words, any interjections, comments made
by the Speaker or Deputy Speaker and continuations within an individual
speech were all parsed together as a single string. As such, the name,
name ID, electorate and party details were only provided for the person
whose turn it was to speak. There were many intricacies in the task of
splitting these speeches in a way that would be generalizable across
sitting days. Details on these are provided later.

Since we are looking at a wide time span of documents, there are many
changes in the way they are formatted. These became apparent as we ran
our script on XML files from earlier sitting days. Some changes are as
subtle as a differently named child node, while others are as extensive
as a different nesting structure. Smaller changes were accounted for as
we became aware of them, and embedded into the code in a way that would
not cause issues for parsing more current Hansards with subtle
differences in formatting. However, as mentioned, more significant
changes in the XML structure of Hansard necessitated the development of
separate scripts as we worked backwards. Further not every sitting day
contains every possible XML element. For example, some days did not have
\texttt{\textless{}subdebate.2\textgreater{}} content, and some days did
not have a Federation Chamber proceeding. To improve the
generalizability of these scripts, if-else statements were embedded
within the code wherever an error might arise due to a missing element.
For example, the entire Federation Chamber block of code is wrapped in
an if-else statement for each script, so that it only executes if what
the code attempts to parse exists in the file.

Once the script ran without error for a few recent years of Hansard, we
continued to work backwards until extensive changes in tree structure
made our script incompatible with parsing earlier XML files. The
earliest sitting day this first script can successfully parse is 14
August 2012. Before developing new scripts to parse earlier Hansard
documents, we prioritized cleaning and finalizing what we had been able
to parse. As such we continued building our script, fixing any problems
we noticed in the resulting datasets such as excess whitespace or
spacing issues, and splitting up any additional sections of the parsed
text onto separate rows where necessary. Specifically, we added a
section of our script to separate out general stage directions. More
information on this separation will be provided in the Stage Directions
section. After completing our first script, it was formatted as a
function which takes a single file name argument and produces one CSV
file containing data on all proceedings from the given sitting day.

\hypertarget{sec-diff}{%
\subsection{Script Differences}\label{sec-diff}}

As mentioned, we developed a total of four scripts to parse the
1998-2022 time frame of Hansard documents. Two main factors motivated us
to create four scripts as opposed to just one, the first being
structural variation in XML over time, and the second being improved
computational efficiency with separate scripts. While all four scripts
use the same general approach to parsing described in the Overview
section and produce the same CSV structure, the first and second scripts
use a different method of data processing than the third and fourth
scripts.

The need for a second script stems from the fact that when established
in 1994, the Federation Chamber was originally named the Main Committee.
The Main Committee was renamed to the Federation Chamber in
mid-2012\textsuperscript{8, ch.~21}. As a result, the child node under
which Federation Chamber proceedings are nested is named
\texttt{\textless{}maincomm.xscript\textgreater{}} in all XML files
prior to 14 August 2012. Having developed our first script based on
Hansard from recent years, all XPath expressions for parsing Federation
Chamber proceedings contain the
\texttt{\textless{}fedchamb.xscript\textgreater{}} specification. To
avoid causing issues in our first script which successfully parses about
10 years of Hansard, we created a second script where we replaced all
occurrences of \texttt{\textless{}fedchamb.xscript\textgreater{}} with
\texttt{\textless{}maincomm.xscript\textgreater{}}. After making this
modification and accounting for other small changes such as timestamp
formatting, this second script successfully parses all Hansard sitting
days from 10 May 2011 to 28 June 2012 (inclusive).

While the modifications needed to develop the second script were
straightforward, this was not the case for our next script. The typical
tree structure of Hansard XMLs spanning from 1998 to March 2011 has an
important difference from that of XMLs released after March 2011,
necessitating many changes to be made in our methodology. In XMLs after
March 2011, which our first two scripts successfully parse, the first
two child nodes of \texttt{\textless{}speech\textgreater{}} are
typically \texttt{\textless{}talk.start\textgreater{}}, and
\texttt{\textless{}talk.text\textgreater{}}. The first child node
contains data on the person whose turn it is to speak, and the second
contains the entire contents of that speech --- including all
interjections, comments, and continuations. After the
\texttt{\textless{}talk.text\textgreater{}} element closes, there are
typically a series of other child nodes which provide a skeleton
structure for how the speech proceedings went in chronological order.
For example, if the speech began, was interrupted by an MP, and then
continued uninterrupted until the end, there would be one
\texttt{\textless{}interjection\textgreater{}} node and one
\texttt{\textless{}continuation\textgreater{}} node following the
\texttt{\textless{}talk.text\textgreater{}} node. These would contain
details on the MP who made each statement, such as their party and
electorate.

In contrast, the speech contents in XMLs from 1998 up to and including
24 March 2011 are nested differently --- there is no
\texttt{\textless{}talk.text\textgreater{}} node. Rather than this
single child node that contains all speech content, statements are
categorized in individual child nodes. This means that unlike our code
for parsing more current Hansards, we cannot specify a single XPath
expression such as
``\texttt{chamber.xscript//debate//speech/talk.text}'' to extract all
speeches, in their entirety, at once. This difference in nesting
structure made many components of our second script unusable for
processing transcripts preceding 10 May 2011, and required us to change
our data processing approach considerably.

\begin{figure}

{\centering \includegraphics[width=3.28125in,height=\textheight]{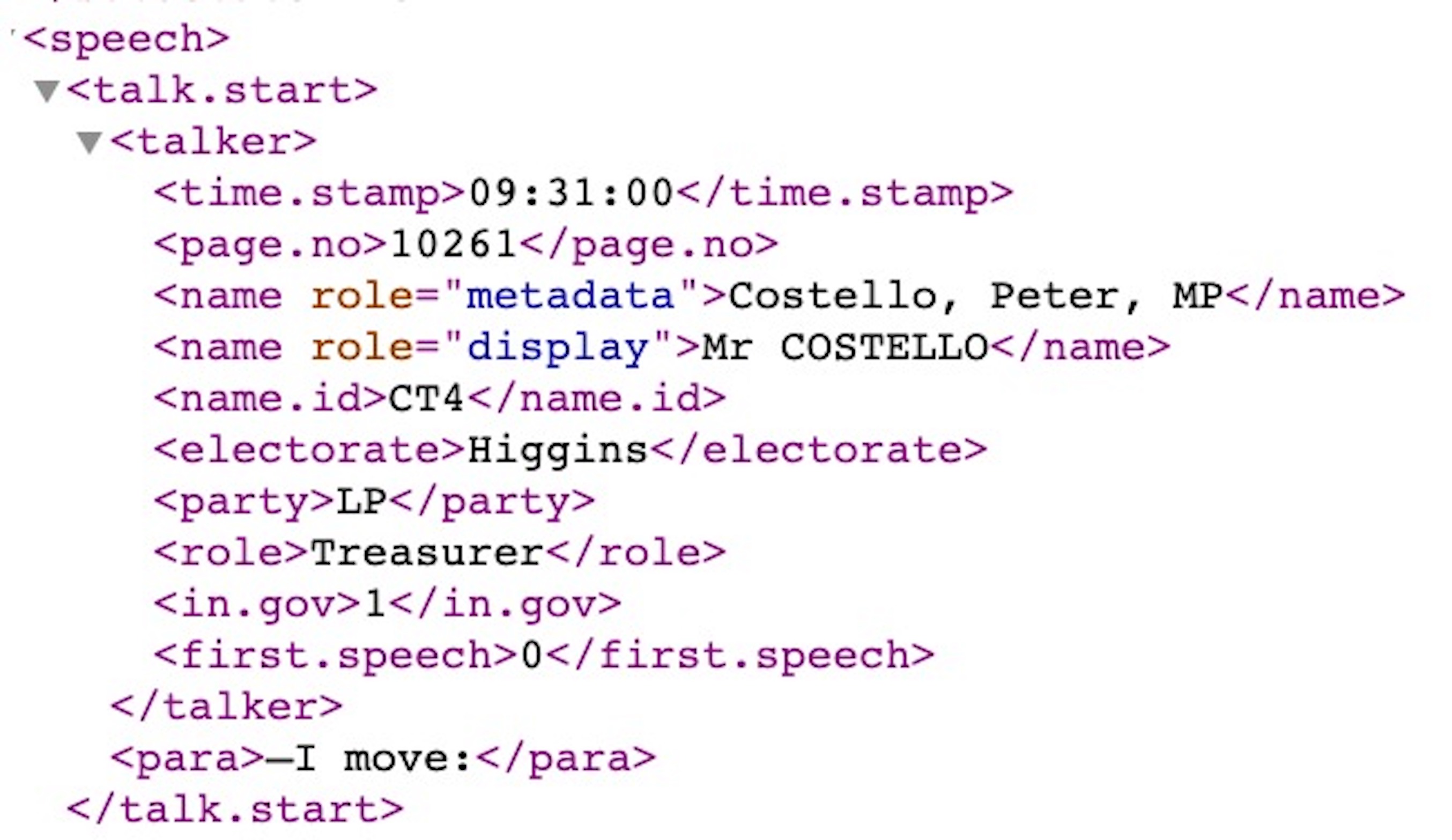}

}

\caption{\label{fig-patternEX}Portion of XML file for Hansard on 12
December 2002}

\end{figure}

Since the earlier Hansard XMLs do not have a
\texttt{\textless{}talk.text\textgreater{}} node, we found that the most
straightforward way to preserve the ordering of statements and to parse
all speech contents at once was to parse from the
\texttt{\textless{}debate\textgreater{}} element directly. The reason we
did not use its \texttt{\textless{}speech\textgreater{}} child node is
because every speech has a unique structure of node children, and this
makes it difficult to write code for data cleaning which is
generalizable across all speeches and sitting days. The challenge with
parsing through the \texttt{\textless{}debate\textgreater{}} element is
that every piece of data stored in that element is parsed as a single
string, including all \texttt{\textless{}talk.start\textgreater{}} data,
and all nested sub-debate data. For example, the
\texttt{\textless{}talker\textgreater{}} data shown in
Figure~\ref{fig-patternEX} would be parsed as a single string preceding
the speech content, like so:

\begin{quote}
\texttt{09:31:0010261Costello,\ Peter,\ MPMr\ COSTELLOCT4HigginsLPTreasurer10}
\end{quote}

This was not isolated to just the beginning of speeches --- details on
individuals interjecting or commenting during speeches were also
captured this way. To separate statements correctly, we collected all of
these patterns using the \texttt{\textless{}talk.start\textgreater{}}
node, and used them to split statements wherever one of these patterns
was found. After separating the statements, we were able to remove these
patterns from the body of text. We also used this method of extracting
and later removing unwanted patterns for other pieces of data which did
not belong to the debate proceedings, such as sub-debate titles.

Once we finalized this new method of processing the data, we proceeded
with data cleaning using the same general approach as in the first two
scripts to produce the same structure of CSV output. We then worked
backwards in time and modified the code as needed for generalizability.
Throughout this process we found a number of transcription errors
present in the XMLs from earlier years. We fixed these manually,
deferring to the official release to ensure the correct information was
filled in. Since there were a number of transcription errors specific to
the 2000s, we chose to create a fourth script for parsing 1998 and 1999.
This allowed us to remove all the code which was needed to resolve
specific transcription errors of the 2000s, to avoid an overly long
script and in turn improving computational efficiency. As such, our
fourth script is essentially the same as the third, with the only
difference being that it has code specific to fixing transcription
errors from 1998 and 1999.

\hypertarget{sec-qa}{%
\subsection{Question Time}\label{sec-qa}}

A key characteristic of the Australian parliamentary system is the
ability for the executive government to be held accountable for their
decisions. One core mechanism by which this is achieved is called
Question Time. This is a period of each sitting day in the Chamber where
MPs can ask ministers two types of questions: questions in writing which
are written in advance, or questions without notice which are asked
verbally in the Chamber and are responded to in real
time\textsuperscript{18}. Questions without notice are included directly
in the \texttt{\textless{}chamber.xscript\textgreater{}} child node,
with sub-child nodes called \texttt{\textless{}question\textgreater{}}
and \texttt{\textless{}answer\textgreater{}} to differentiate the two.
Questions in writing, however, are embedded in their own child node
called \texttt{\textless{}answers.to.questions\textgreater{}} at the end
of the XML file.

Our approach to parse the
\texttt{\textless{}chamber.xscript\textgreater{}} speeches used in all
four scripts meant that all questions without notice content was already
parsed in order. For the first two scripts, questions and answers were
already separated onto their own rows. For the third and fourth scripts,
just as we did with the rest of the speech content, we used those
patterns of data preceding the text to separate questions and answers.
Finally, since questions in writing exist in their own child node we
were able to use the same parsing method for all scripts, which was to
extract all question and answer elements from the
\texttt{\textless{}answers.to.questions\textgreater{}} child node.

We then added binary flags to differentiate between questions and
answers. To do this in the first and second scripts, we separately
re-parsed question and answer content using the XPath expressions
\texttt{"chamber.xscript//question"} and
\texttt{"chamber.xscript//answer"}, added the correct question and
answer flags accordingly, and then added those flags back to the main
dataframe based on exact text matches. For the third and fourth scripts,
we made use of the fact that the patterns preceding text transcribed
under a question node were stored separately from those transcribed
under an answer node. As a result, we could readily use those patterns
to flag questions and answers correctly based on which list of patterns
it belonged to. Sometimes, we identified questions which were
incorrectly transcribed under an answer node and vice-versa, in which
cases we manually corrected the question and answer flags. For instance,
we check for any statements flagged as questions which include the
phrase ``has provided the following answer to the honourable member's
question'', in which case we re-code that statement as an answer. It is
important to note, however, that because we identified and corrected for
these transcription errors manually as we discovered them, additional
flagging errors may exist which we did not catch. As such, users may
identify and should be wary of occasional incorrectly flagged questions
or answers in the data.

The next step was to merge Question Time contents with all the debate
speech. As mentioned, our method of parsing meant that everything was
already in order, so we did not have to perform any additional merging
for questions without notice content. For questions in writing, merging
this content was also straightforward due to the fact that it is always
at the end of Hansard. This means that we could bind question in writing
rows to the bottom of the main dataframe. This approach was used for all
four scripts.

\hypertarget{sec-interject}{%
\subsection{Interjections}\label{sec-interject}}

As mentioned, the text was structured and parsed in such a way that
various interjections and comments which happened during a speech were
not separated onto individual rows. This was the case across the entire
time frame of documents. We will first discuss the methodology employed
to split interjections in the first and second scripts, as it informed
our approach for the third and fourth scripts.

Below is an example of part of a speech we would need to split,
extracted from Hansard on 30 November 2021, where Bert van Manen is
interrupted by the Speaker who states that the time for members'
statements has concluded.

\begin{quote}
``Mr VAN MANEN (Forde---Chief Government Whip) (13:59): It's a great
pleasure to share with the House that Windaroo Valley State High School
has qualified for the finals of the Australian Space Design Competition,
to begin in January next year. The competition is regarded as the
premier STEM competition for high school students and is recognised by
universities around the country. The students are required to respond to
industry-level engineering and requests for tender for design and---The
SPEAKER: Order! In accordance with standing order 43, the time for
members' statements has concluded.''
\end{quote}

We want each statement on its own row with the correct name, name ID,
electorate and party information on the individual speaking. We
approached this task in a number of steps.

Once all parsed text from the XML was merged into one dataframe called
\texttt{main}, our first step was to add a \texttt{"speech\_no"}
variable. This was done to keep track of which speech each interjection,
comment, or continuation belonged to as we separated these components
onto their own rows.

The next step was to extract all the names and titles preceding these
interjections, comments and continuations. This would enable us to then
separate the speeches in the correct places using these names and titles
in combination with regular expressions, which are patterns of
characters that can be used to search bodies of text. We completed this
extraction process with a few intermediate steps, due to the large
number of name styles and interjection types that had to be accounted
for, each requiring their own unique regular expression format.

\begin{figure}

{\centering \includegraphics[width=3.59375in,height=\textheight]{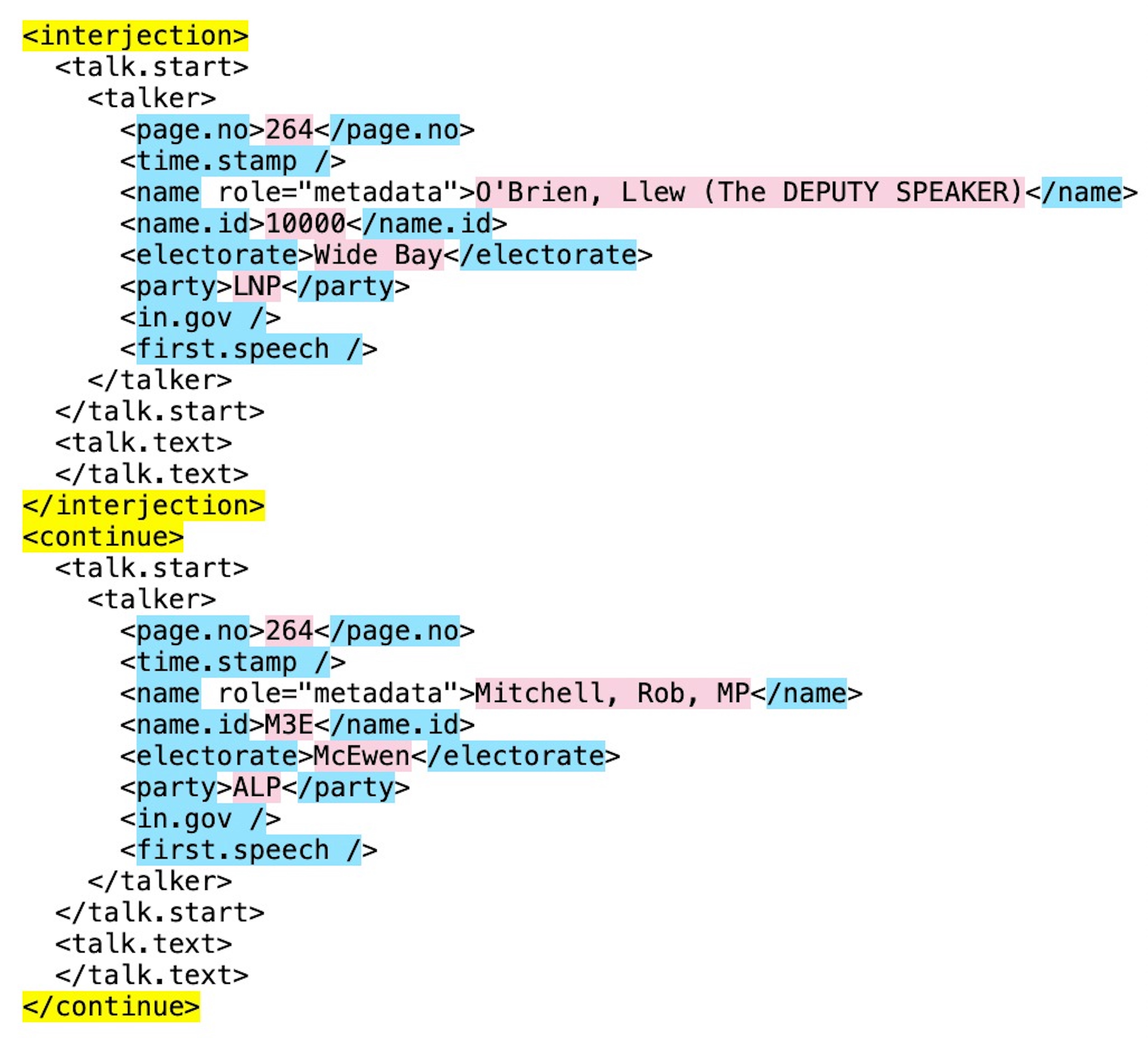}

}

\caption{\label{fig-xml2}Snapshot of XML structure with interjection and
continuation from 03 February 2021 Hansard}

\end{figure}

As mentioned earlier, more recent years of Hansard XMLs contain a series
of child nodes which exist to capture the structure of interruptions in
that speech. Figure~\ref{fig-xml2} provides an example of this, where
the speech was interrupted by a comment from the Deputy Speaker, and
then the MP continued their speech. Looking at the element names
highlighted in blue, these child nodes do not contain the actual text
for the interjection or continuation --- this text is embedded within
the speech above it. However, as shown by the content highlighted in
pink in Figure~\ref{fig-xml2}, we were able to extract useful details on
the individual interjecting which we could use later. Making use of this
structure, we extracted names and information of all individuals that
were categorized within the XML as interjections. We stored this as a
dataframe called \texttt{"interject"}. We decided not to include this
data in our final database, as it is embedded in our resulting datasets
which have a flag for interjections.

We then created lists using both the \texttt{interject} and
\texttt{main} dataframes to capture all the names of individuals who
spoke that day. We added the names of all MPs in a number of unique
formats, due to the frequent variation in how names are transcribed in
Hansard. When an MP interjects or continues a speech, the usual form of
their name is a title followed by their first name or first initial
and/or last name. There is also variation in the capitalization of these
names. Sometimes when someone's first name is included, only their last
name is capitalized, while sometimes their full name is capitalized, or
other times neither are capitalized. Another source of variation is in
individuals with more than one first name, as sometimes only their
initial first name is written, while other times their entire first name
is written. Additionally, some surnames have punctuation, and some
surnames have specific capitalization such as ``McCormack'', where even
in full capitalization, the first ``c'' remains lower case. This
variation demands careful consideration when writing regular expression
patterns. In these lists we also accounted for any general interjection
statements that were not attributed to an individual, such as ``An
opposition member interjecting-''.

Having these lists enabled us to extract the names of MPs and their
associated titles as they exist in the text, by searching for exact
matches with regular expression patterns. We then used these extracted
names to split all the speeches, using regular expressions with
lookarounds. A lookaround can be added to a regular expression pattern
to enhance the specificity of matches. These were used to ensure that
the text was not being split in the wrong places, such as places where
MPs were being named in the statement of another MP.

Once all interjections, comments and continuations were split onto their
own rows using the lists we created, we did one final check for any
additional names that were not captured in these lists. We searched for
any remaining name matches in speech bodies with general regular
expressions and lookarounds, and separated text using those matches when
found.

\begin{figure}

{\centering \includegraphics[width=4.66667in,height=\textheight]{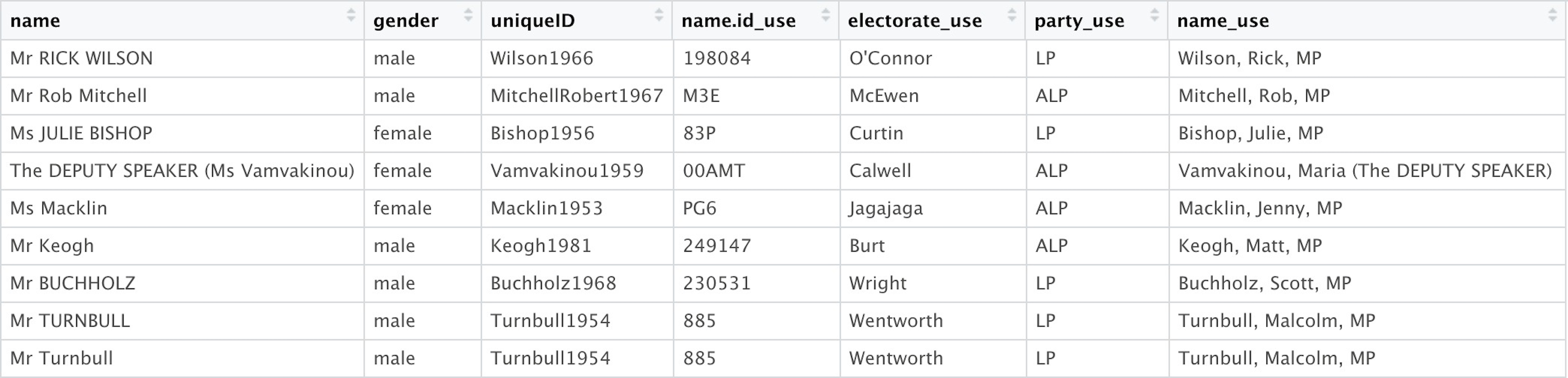}

}

\caption{\label{fig-lookup}First 10 rows of the lookup table from 19
October 2017 Hansard processing}

\end{figure}

We then added an order variable to the dataset based on row number, to
keep track of the order in which everything was said. The next step was
to fill the name, name ID, electorate and party variables with the
correct data for each row. We also wanted to add the gender and unique
identifier for each individual as found in the
\texttt{AustralianPoliticians} package. To do so, we created a lookup
table, which contained the unique incomplete form in which the name was
transcribed, and the corresponding full name, name ID, electorate,
party, gender, and unique ID for that individual.
Figure~\ref{fig-lookup} provides an example of this. We used the main
dataset from the \texttt{AustralianPoliticians} package in the creation
of each lookup table\textsuperscript{9}.

\begin{figure}

{\centering \includegraphics{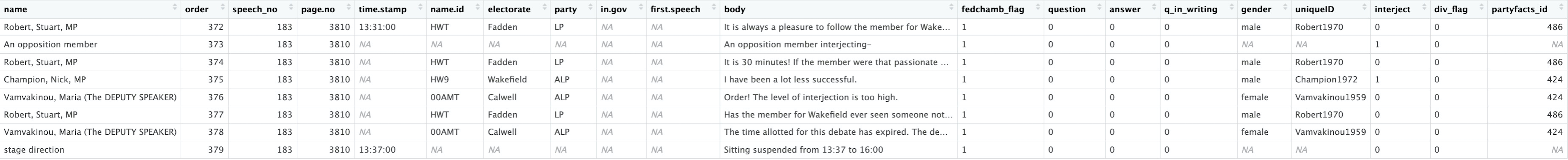}

}

\caption{\label{fig-interject}Example of speech with interjections from
21 November 2016 Hansard}

\end{figure}

Next, we merged our main dataframe with the lookup table to replace any
incomplete names with their full names, and to fill in any gaps with
available name ID, electorate, party, gender, and unique ID information.
Finally, we were able to add a flag for interjections. Grouping our data
by the speech number, we defined an interjection as a statement made by
anyone who is not the Speaker, the Deputy Speaker, or the MP whose turn
it was to speak. Figure~\ref{fig-interject} provides an example of a
Federation Chamber proceeding with interjections. Statements made by the
MP whose turn it was to speak, or by the Deputy Speaker Maria
Vamvakinou, are not flagged as interjections.

Having developed a successful methodology for splitting interjections,
we used this to inform our general approach in the third and fourth
scripts. However, the difference in data cleaning used in these scripts
necessitated some departure from the original methodology. As discussed
earlier, we used string patterns extracted from
\texttt{\textless{}talk.start\textgreater{}} nodes to separate speeches.
As evident in Figure~\ref{fig-xml2},
\texttt{\textless{}talk.start\textgreater{}} nodes are nested within
\texttt{\textless{}interjection\textgreater{}} nodes, meaning that the
patterns of data from interjection statements were separated out in the
process. This meant that we did not need to create lists of names and
titles for which to search in the text as we did before. However, we
used the same list of general interjection statements on which to
separate as was used in the first two scripts. We then did an additional
check for statements that may have not been separated due to how they
were embedded in the XML, and separated those out where needed. In
particular, while most statements were categorized in their own child
node and hence captured through pattern-based separation, some were not
individually categorized, and had to be split manually in this step.

\begin{figure}

{\centering \includegraphics{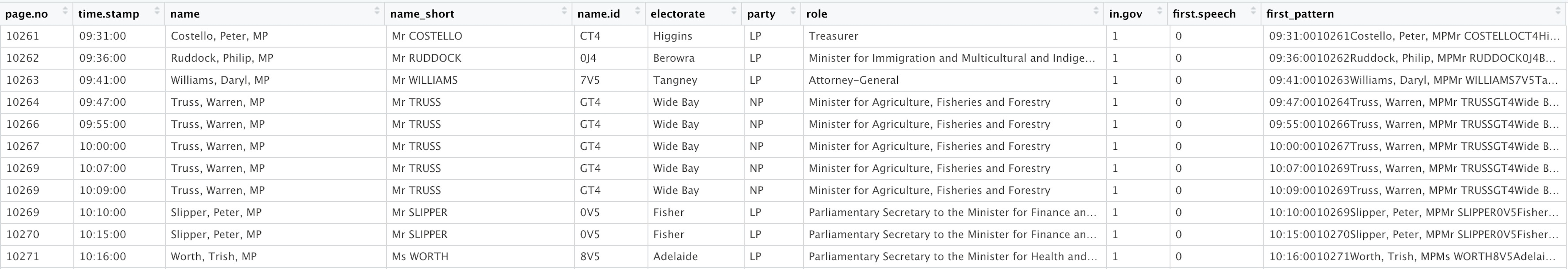}

}

\caption{\label{fig-patterns}10 rows of the pattern lookup table from 12
December 2012 Hansard processing}

\end{figure}

We then proceeded to clean up speeches and fill in correct details on
the MP speaking. While we used the same lookup table approach as before,
we did so in combination with another means of filling in these details.
The patterns parsed from \texttt{\textless{}talk.start\textgreater{}}
nodes contain important data on the MP making each statement. As such,
we could extract those data associated with each pattern by parsing one
element inward, using the XPath expression \texttt{"talk.start/talker"}.
We created a pattern lookup table with these data, and merged it with
the main Hansard dataframe by the first pattern detected in each
statement. Figure~\ref{fig-patterns} provides an example of that lookup
table. This approach enabled us to fill in missing data on each MP
speaking using data extracted directly from the XML. Finally, we then
used the \texttt{AustralianPoliticians} dataset to fill in other missing
data, and flagged for interjections in the same manner as before.

\hypertarget{sec-stage}{%
\subsection{Stage Directions}\label{sec-stage}}

When building our first scripts, one of the final components needed was
to separate general stage directions out from statements made by MPs.
Stage directions are general statements included in the transcript to
document happenings in parliament. Examples of stage directions are
``Bill read a second time'', ``Question agreed to'', or ``Debate
adjourned''. It was unclear to us from the XML and PDF who exactly these
statements were attributed to. For further clarification, we watched
portions of the video recording for some sitting days, and noticed that
while these statements are documented in Hansard, they are not
explicitly stated in parliament. For example, when the Deputy Speaker
says ``The question is that the bill be now read a second time'', MPs
vote, and if the majority is in favour, they proceed reading the bill
the second time. This vote and second reading is not explicitly
transcribed, rather what is written is: ``Question agreed to. Bill read
a second time''. For this reason, we filled the name variable for these
statements with ``stage direction''. Stage directions were not flagged
as interjections. These stage directions are not defined differently
from the regular debate speech in the XML, meaning we had to manually
create a list of stage directions to separate out of the speeches. We
built this list of stage directions as we worked backwards in parsing
Hansard, and took the same approach across all four scripts. Despite our
best efforts to capture all stage directions in this list, because it
had to be built manually, users should be aware that it is possible that
some stage directions were not separated onto their own rows in the
process. Further, it is important to note that because they are not
spoken aloud, stage directions do not represent speech components in the
same way that all other components such as interjections and
continuations do. However, they provide valuable information about the
inner-workings and structure of parliamentary events. If the focus of
the user's research does not require stage directions, then this can be
removed by filtering out observations in \texttt{name} equal to
``business start'' or ``stage direction''.

\hypertarget{filling-missing-details}{%
\subsection{Filling Missing Details}\label{filling-missing-details}}

While we did our best to maximize the completeness of the files in our
database as they were processed in the initial four scripts, there were
still a number of rows in which details on the person speaking were
missing, or the name transcribed for that individual was in a short form
(e.g.~``Mr Abbott'' instead of ``Abbott, Tony, MP''). This was a
particularly frequent occurrence for sitting days where an MP spoke
whose surname was shared by any other past or present MP, and automated
filling of their details using data from the
\texttt{AustralianPoliticians} package was avoided to prevent any
incorrect detail attribution. In an effort to improve as many of these
as possible, we developed a script which identifies short-form names
belonging to people with common surnames in each CSV, looks for the full
version of that individuals name if available in that same CSV file, and
replaces the short-form name with the full name, and fills the rest of
the MP details in accordingly with data from the
\texttt{AustralianPoliticians} package. This script does the same for
anyone who does have a unique surname but is still missing the full name
form or any gender, unique ID, name ID, party or electorate details.
Each file in our database passed through this script after being
created, to ensure it is as complete as possible.

Due to the fact that the names of MPs with common surnames were not all
in their complete form when we flagged for interjections the first time,
it was possible that the name of the MP whose turn it was to speak was
transcribed in different forms within their speech. For example,
``Smith, Tony, MP'' at the start and then ``Mr Smith'' later on in the
speech. By the nature of how we flagged for interjections, this means
that rows where the short form like ``Mr Smith'' is the name would be
flagged as an interjection, which is incorrect. To fix this, we
re-flagged interjections using the same definition as before, once all
names were filled in with this script.

\hypertarget{debate-topics}{%
\subsection{Debate Topics}\label{debate-topics}}

To enhance the range of research questions which can be explored with
our data, we have created a supplementary file containing debate topics
and their corresponding page numbers for each sitting day in our
database. To extract these data, we wrote a script to parse debate and
sub-debate 1 information elements from each XML file in chronological
order using the XPath expression
``\texttt{//debate/debateinfo\ \textbar{}\ //subdebate.1/subdebateinfo}'',
and added a date variable with the date of each sitting day. Note that
in some cases there were multiple page number child nodes for the same
debate or sub-debate title, likely due to manual transcription error.
Upon manual inspection, we found that most often, the second page number
node contained the same page number as the first node, and sometimes the
second node contained a repeated debate title or a timestamp. Therefore,
we took the first available page number child-node for each debate or
sub-debate node to be the one we included in our dataset. After
summarizing, these topics can be added to the main text by joining.
Example code is provided in the README.

\hypertarget{divisions}{%
\subsection{Divisions}\label{divisions}}

Another fundamental component of parliamentary proceedings is voting. In
the House, when a question arises such as ``The question is that the
amendment be agreed to'', Members are asked to cast their vote either in
the affirmative or in the negative, and the majority of votes as judged
by the Speaker determines the outcome\textsuperscript{8}. Further,
Members can vote in an unofficial arrangement called pairs, which ``can
be used to enable a Member on one side of the House to be absent for any
votes when a Member from the other side is to be absent at the same time
or when, by agreement, a Member abstains from
voting''\textsuperscript{8}. If the result as determined by the Speaker
is challenged by more than one MP, this leads to a division of the House
in which the question is re-stated and Members must move to the left or
right of their chair depending on how they vote, so that the votes can
be re-counted and recorded\textsuperscript{8}.

In the Hansard XML files, divisions data are structured outside of the
\texttt{\textless{}speech\textgreater{}} content in their own
\texttt{\textless{}division\textgreater{}} nodes that contain the voting
data and division result. Since we focus primarily on the spoken
\texttt{\textless{}speech\textgreater{}} Hansard content, our parsing
scripts do not necessarily capture all divisions data from House
proceedings. Our approach to parsing Hansard in the third and fourth
scripts described in the Script Differences section naturally allowed
for much of the divisions data to be added to our resulting files for
1998 to March 2011, however the parsing scripts used for May 2011 to
September 2022 Hansard did not. To supplement our database and in an
effort to fill this divisions data gap, we created an additional file
containing all divisions data nested under the XPath
``\texttt{//chamber.xscript//division}'' from the Hansard files in our
time frame. To produce this data file, for each Hansard XML we parsed
the \texttt{\textless{}division.header\textgreater{}},
\texttt{\textless{}division.data\textgreater{}}, and
\texttt{\textless{}division.result\textgreater{}} child-nodes where they
existed, extracted any timestamps where available, and did any
additional data cleaning as necessary. We used a series of if-else
statements in this script to account for variation in the structure of
the \texttt{\textless{}division\textgreater{}} node over time. Finally,
we then added a date variable to distinguish between sitting days.

\hypertarget{data-records}{%
\section{Data Records}\label{data-records}}

Our database is available in both CSV and parquet formats. Both CSV and
parquet are open standards. We provide both because while CSVs are
commonly used and can be manually inspected, parquet files are typically
smaller and preserve class. Our database covers all sitting days of the
House of Representatives from 02 March 1998 to 08 September 2022 where
an XML transcript is available, so there are 1,532 individual sitting
day files for each format. Additionally, there is a single corpus file
in both CSV and parquet forms containing the data from all sitting days,
with a date variable added to allow for distinction and filtering of
individual sitting days. There is also a CSV and parquet file containing
all parsed debate topics. All data records are available on the
general-purpose repository Zenodo, at
\url{https://doi.org/10.5281/zenodo.7336075}\textsuperscript{19}. For
each Hansard data file, that is, the corpus and the individual sitting
day files, each row contains an individual statement, with details on
the individual speaking. For general statements transcribed as made by
``Honourable members'' for example, these variables cannot be specified.
Table~\ref{tbl-vars} provides an overview of each variable found in the
Hansard data files in the database.

\hypertarget{tbl-vars}{}
\begin{table}[H]
\caption{\label{tbl-vars}Summary and description of variables in our database }\tabularnewline

\centering
\begin{tabular}{>{\raggedright\arraybackslash}p{8em}>{\raggedright\arraybackslash}p{30em}}
\toprule
Variable & Description\\
\midrule
name & Name of speaker\\
order & Individual row numbering, based on our data processing and text separation method\\
speech\_no & Index associated with each speech made on the given sitting day, including all statements, comments and interjections made in the duration of that speech\\
page.no & Page number statement can be found on in official Hansard\\
time.stamp & Time of statement\\
\addlinespace
name.id & Unique member identification code, based on the Parliamentary Handbook\\
electorate & Speaking member's electorate\\
party & Speaking member's party\\
in.gov & Flag for in government (1 if in government, 0 otherwise)\\
first.speech & Flag for first speech (1 if first speech, 0 otherwise)\\
\addlinespace
body & Statement text\\
fedchamb\_flag & Flag for Federation Chamber (1 if Federation Chamber, 0 if Chamber)\\
question & Flag for question (1 if question, 0 otherwise)\\
answer & Flag for answer (1 if answer, 0 otherwise)\\
q\_in\_writing & Flag for question in writing (1 if question in writing, 0 otherwise)\\
\addlinespace
gender & Gender of speaker\\
uniqueID & Unique identifier of speaker\\
interject & Flag for interjection (1 if statement is an interjection, 0 otherwise)\\
div\_flag & Flag for division (1 if division, 0 otherwise)\\
partyfacts\_id & PartyFacts identification number, based on linked parties data from the Party Facts project\\
\bottomrule
\end{tabular}
\end{table}

The \texttt{name}, \texttt{page.no}, \texttt{time.stamp},
\texttt{name.id}, \texttt{electorate}, \texttt{party}, \texttt{in.gov},
\texttt{first.speech}, and \texttt{body} variables all came directly
from the XML contents. In addition to these variables, we added a number
of flags to enable easy filtering of statements. For example, adding the
\texttt{fedchamb\_flag} provides a clear distinction between the
proceedings of the Chamber with those of the Federation Chamber. The
\texttt{question}, \texttt{answer}, and \texttt{q\_in\_writing} flags
were added to identify statements belonging to Question Time, and the
nature of these statements. We also flagged for interjections
(\texttt{interject}), and the \texttt{div\_flag} variable was added to
flag those rows where ``The House divided.'' was detected in the body
variable. The \texttt{gender} and \texttt{uniqueID} variables were added
based on the main dataset from the \texttt{AustralianPoliticians}
package, and the \texttt{partyfacts\_id} variable was added using code
and data provided by the Party Facts Project website. Note that in
accordance with the code provided on the Party Facts downloads page
(https://partyfacts.herokuapp.com/download/), we used only the core
datasets maintained by the Party Facts project, which are the Manifesto
Project and ParlGov. Details on these datasets can be found on the Party
Facts datasets documentation page
(https://partyfacts.herokuapp.com/documentation/datasets/). Details on
the usage of \texttt{uniqueID} and \texttt{partyfacts\_id} will be
provided in the Usage Notes to follow. Further, the \texttt{speech\_no}
variable allows us to keep track of the speech number that each
statement and interjection belongs to. Having the speech number variable
offers an easy way to group statements by speech or isolate specific
speeches of interest. Lastly, the \texttt{order} variable was added to
maintain the order of proceedings, after all individual statements were
separated onto their own rows.

As mentioned, in addition to the Hansard data described above, our
database also contains a CSV and parquet file with the parsed debate
topics. This file is called \texttt{all\_debate\_topics}, and contains a
\texttt{date} variable specifying the sitting day, an
\texttt{item\_index} variable to specify the order of proceedings
(i.e.~the order in which these topics were discussed), a \texttt{title}
variable containing the debate or sub-debate title contents, and a
\texttt{page.no} variable specifying on which page that title was
recorded to be found in the official Hansard PDF.

There is also a CSV file in our database called
\texttt{PartyFacts\_map.csv}. This file was created using
\texttt{AustralianPoliticians} data, data downloaded from the Party
Facts project, and our own Hansard data \texttt{party} variable. As
there are some party name and abbreviation spelling inconsistencies
across these sources, creating this dataframe allowed us to ensure
correct merging of PartyFacts ID numbers with their associated party, in
accordance with the party abbreviation spelling transcribed in our
Hansard data. Further, we included the corresponding party abbreviation
and full name spelling found in the \texttt{AustralianPoliticians}
package for completeness.

\hypertarget{tbl-partyfacts-vars}{}
\begin{table}[H]
\caption{\label{tbl-partyfacts-vars}Summary and description of variables in our Party Facts mapping file }\tabularnewline

\centering
\begin{tabular}{>{\raggedright\arraybackslash}p{10em}>{\raggedright\arraybackslash}p{25em}}
\toprule
Variable & Description\\
\midrule
partyfacts\_id & PartyFacts identification number, based on linked parties data from the Party Facts project\\
party\_abb\_hansard & Party abbreviation as transcribed in the Hansard data\\
party\_abb\_auspol & Party abbreviation as recorded in the AustralianPoliticians R package data\\
party\_name\_auspol & Party full name as recorded in the AustralianPoliticians R package data\\
\bottomrule
\end{tabular}
\end{table}

Finally, our database contains a file with all parsed divisions data in
our timeframe. This is called \texttt{division\_data}, and is available
in RDA form and in parquet form. The reason it is not available in CSV
form is because three of the variables (\texttt{names\_AYES},
\texttt{names\_NOES}, and \texttt{names\_PAIRS}) are lists, which are
not supported by the CSV rectangular data structure. The variables found
in these data are summarized below, in Table~\ref{tbl-divisions}.

\hypertarget{tbl-divisions}{}
\begin{table}[H]
\caption{\label{tbl-divisions}Summary and description of variables in our divisions data file }\tabularnewline

\centering
\begin{tabular}{ll}
\toprule
Variable & Description\\
\midrule
date & Sitting day\\
div\_num & Index for each division parsed for that sitting day\\
time.stamp & Timestamp associated with each division\\
num.votes\_AYES & Number of votes in the affirmative\\
num.votes\_NOES & Number of votes in the negative\\
\addlinespace
num.votes\_PAIRS & Number of votes in pairs\\
names\_AYES & List of the names of Members who voted in the affirmative\\
names\_NOES & List of the names of Members who voted in the negative\\
names\_PAIRS & List of the names of Members who votes in pairs\\
result & Voting result\\
\bottomrule
\end{tabular}
\end{table}

\hypertarget{technical-validation}{%
\section{Technical Validation}\label{technical-validation}}

We developed a script to perform automated tests on each file in our
database, to enhance its quality and consistency. Our first test
validates that the date specified in each file name matches the date
specified in its corresponding XML session header. This XML component
can be seen in Figure~\ref{fig-xml1}, where the first child node of the
\texttt{\textless{}session.header\textgreater{}} element is the date.
Every file passed this test, and we detected one discrepancy in an XML
file from 03 June 2009, where its session header contained the wrong
date. We validated that our file name and date was correct by checking
the official PDF release from that sitting day.

The second test is designed to test for duplication errors in the data,
by checking whether two lines that immediately follow each other have
the same body (i.e.~spoken content). This test detected 131 dates on
which duplicate statements were made, and one immediately follows the
other. Note that this test does not account for who is making each
statement, meaning one MP repeating the words of another MP would be
picked up in this test as well. We checked a sample of 40\% of these
duplicates, and manually validated that they are all repeated statements
that do exist and are transcribed closely together in that day's XML
file, and by our method should be parsed such that one of these
statements is immediately followed by the other.

When an MP runs out of allotted time for their speech, Hansard editors
transcribe ``(Time expired)'' after their final word. As a means of
checking that we have separated speeches out correctly, our third test
checks that when the phrase ``(Time expired)'' exists in a body of text,
it exists at the very end. When this is not the case, we know that we
have missed the separation of the next statement onto its own row, and
could fix this accordingly.

The fourth test is designed to detect issues in timestamp formatting in
the data, by detecting all timestamps in our database which do not match
the correct ``HH:MM:SS'' format. We found a total of 88 incorrectly
formatted timestamps, with common issues such as the ``HH'' or ``MM''
component recorded as ``NaN'' (e.g.~``NaN:28:00'' or ``09:NaN:00''), as
well as timestamps with a third digit in the minute component
(e.g.~``09:497:00'' or ``13:445:00''). We took a random sample of 25\%
of these incorrectly formatted timestamps, and manually checked whether
or not they were transcribed as such in the original XML Hansard
transcript. We found that every improperly formatted timestamp in our
random sample was in fact transcribed as such in its original XML file,
meaning these bugs are the result of a transcription error, rather than
an error resulting from our data parsing or cleaning code. As such, we
have left these timestamps in their originally transcribed format in our
database.

The remaining tests focus on the MPs present on each sitting day. Our
fifth test checks that there is one unique party and electorate
attributed to each individual on each sitting day. As we parsed Hansard
further back in time, we found a number of cases where an individual was
associated with the wrong electorate or party due to transcription
errors. When we found these data errors we corrected them based on the
official release. This test provides us with an automated way to catch
these errors and correct them at scale.

Next, we test that the unique name identification code attributed to
each individual is found in the Australian Parliamentary Handbook. We do
so using the \texttt{ausPH} package. This test serves as another means
to correct for transcription errors, this time in the case of name IDs.
We found and corrected for a number of common name ID transcription
errors detected by this test, such as a capital letter ``O'' in place of
a zero.

Our seventh test checks that on any given sitting day, the individuals
identified are alive. To do so, we utilized the main dataset from the
\texttt{AustralianPoliticians} package which contains the birth and
where applicable death dates for every politician. This test confirmed
that all MPs who are detected to speak on each sitting day are not
deceased.

Finally, our eighth test validates that all individuals speaking are MPs
on that particular day. We use the \texttt{mps} dataset from the
\texttt{AustralianPoliticians} package which has the dates when each MP
was in parliament. Using these dates, we check that each person speaking
on each sitting day is in fact an MP on that day.

\hypertarget{summary-statistics}{%
\subsection{Summary Statistics}\label{summary-statistics}}

\begin{figure}

{\centering \includegraphics[width=4.58333in,height=\textheight]{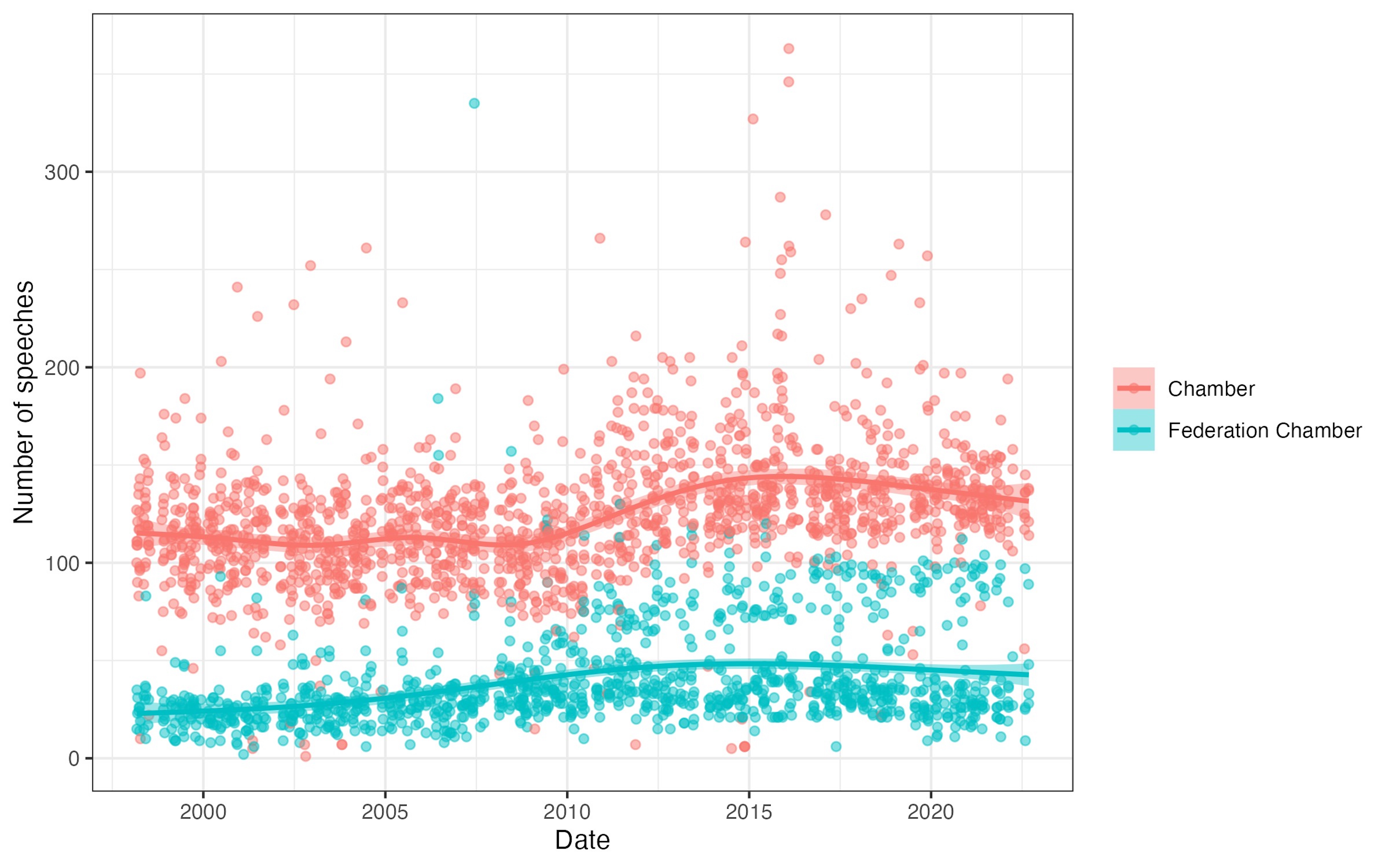}

}

\caption{\label{fig-dailyspeeches}Number of speeches made on each
sitting day in the House of Representatives}

\end{figure}

To further explore the quality of our data and detect any unexpected or
distinct trends, we generated a number of summary statistics using the
complete Hansard corpus. First, we looked at the number of speeches made
each day, disaggregated by Chamber and Federation Chamber debate venues.
As shown in Figure~\ref{fig-dailyspeeches}, there are consistently more
speeches made in the Chamber proceedings than in the Federation Chamber,
which aligns with the wider scope of business covered in the Chamber.
Further, in both debate venues, the number of speeches each day appears
to have increased slightly over time.

\begin{figure}

{\centering \includegraphics[width=4.58333in,height=\textheight]{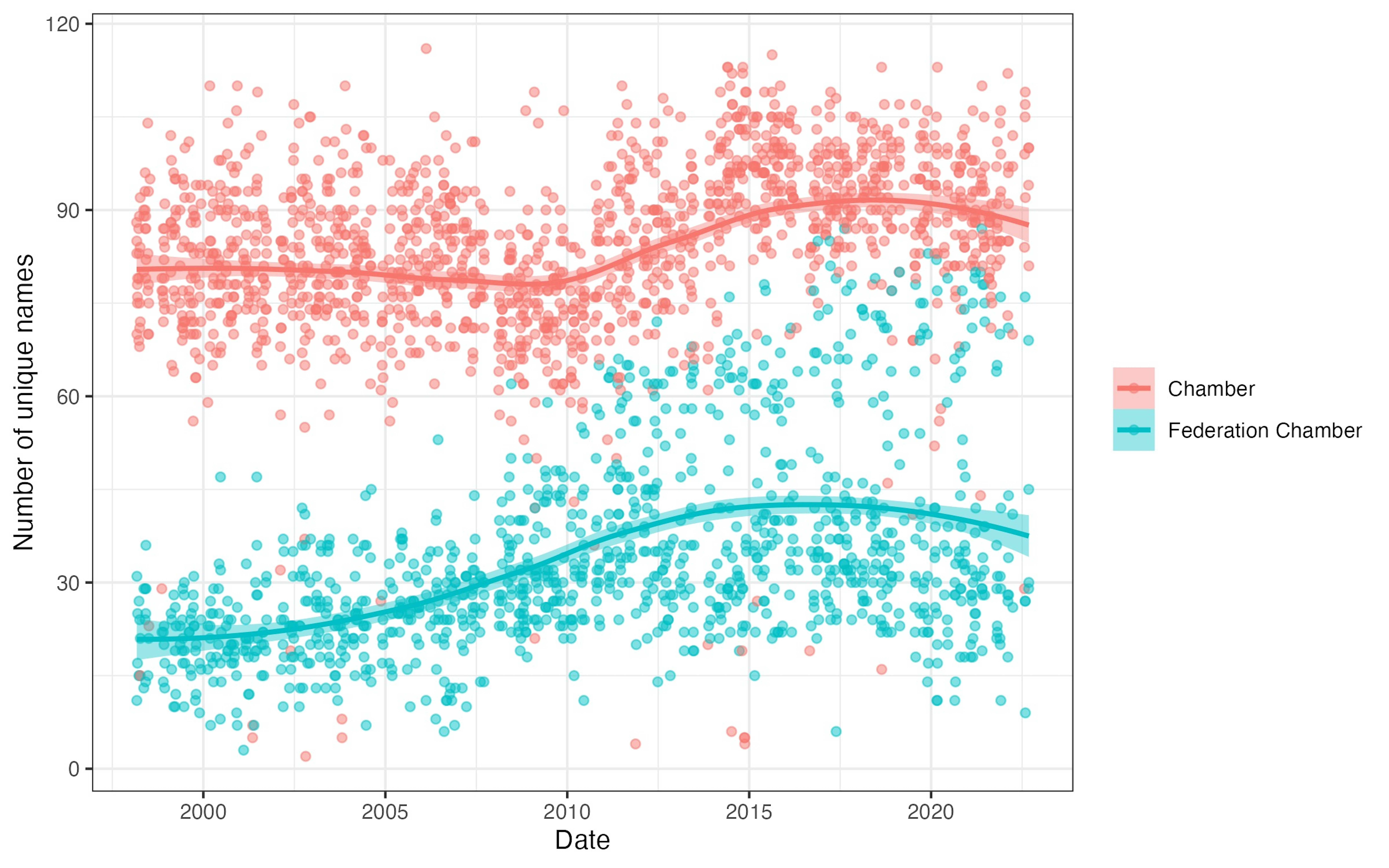}

}

\caption{\label{fig-names}Daily number of unique names detected in our
Hansard data}

\end{figure}

Next, we explored the number of unique names detected on each sitting
day disaggregated by debate venue, visualized in Figure~\ref{fig-names}.
As expected, there are generally more individual MPs detected to be
present in the Chamber proceedings in our data than in the Federation
Chamber proceedings. Further, in both venues, the number of individuals
detected per sitting day appears to have increased since around
2008-2009, with a maximum of 116 and 87 unique names detected for the
Chamber and Federation Chamber, respectively. Across the time frame of
our data, there was a daily average of 84 unique names detected in the
Chamber, and 34 in the Federation Chamber. These observations are in
accordance with the official number of Members belonging to the House of
Representatives, which was 148 in 1998, 150 in 2001, and then increased
to 151 in the 2019 general election\textsuperscript{8}.

We then computed the total number of speeches made in our database by
political party, summarized in Table~\ref{tbl-speeches_by_party}. We
found that the Australian Labor Party made the most speeches overall,
followed by the Liberal Party of Australia, and the National Party of
Australia. These rankings are unsurprising, as those parties are the
three main political parties in the Australian House of
Representatives\textsuperscript{20}.

\hypertarget{tbl-speeches_by_party}{}
\begin{table}[H]
\caption{\label{tbl-speeches_by_party}Total number of speeches made in our database by political party }\tabularnewline

\centering
\begin{tabular}{llr}
\toprule
Party Name & Party Abbreviation & Total \# of Speeches\\
\midrule
Australian Labor Party & ALP & 112,268\\
Liberal Party of Australia & LIB & 106,139\\
National Party of Australia & NPA & 20,244\\
Independent & IND & 5,357\\
Australian Greens & GRN & 1,446\\
\addlinespace
Country Liberal Party (Northern Territory) & CLP & 665\\
Centre Alliance & CA & 638\\
Katters Australian Party & KAP & 263\\
Nick Xenophon Team & NXT & 186\\
National Party of Australia (WA) & NATS WA & 68\\
\addlinespace
Palmer United Party & PUP & 64\\
United Australia Party & UAP & 43\\
\bottomrule
\end{tabular}
\end{table}

\hypertarget{sec-usage}{%
\section{Usage Notes}\label{sec-usage}}

To enhance the usability of our database, we added a \texttt{uniqueID}
variable to each file. This serves as a unique identifier for each
speaking MP, and comes from the \texttt{uniqueID} variable present
within data from both the \texttt{AustralianPoliticians} R
package,\textsuperscript{9} and the \texttt{AustralianElections}
(https://github.com/RohanAlexander/AustralianElections) R package, which
was created by Rohan Alexander. By including this variable, one can
integrate our data records with those available in these two packages.
Similarly, we added the \texttt{partyfacts\_id} variable to our data
which allows users to link our data to external political party data
sources with this unique identifier.

Further, the \texttt{name.id} variable found in each file is another
unique identifier for each MP. This variable was parsed directly from
the Hansard XML files, and can be found in the Australian Parliamentary
Handbook. As such, our data records can be integrated with those from
the \texttt{ausPH} package which provides datasets for contents of the
Australian Parliamentary Handbook. This will allow for convenient
extraction of further details on each MP in a tidy, ready to analyze
format.

Finally, the README file on our GitHub repository and on Zenodo contains
example code on how to read in a file from our database, for both CSV
and parquet formats. Should a user wish to use the single Hansard corpus
rather than individual sitting day files, we provide example code on how
to read it in, filter for those sitting days of interest, and split them
into distinct data frames. We have also added example code for filtering
out stage directions, and updating the \texttt{order} variable to
reflect the ordering of remaining rows, should the user wish to remove
stage directions from their analysis. Lastly, the README contains
example code showing users how to merge the debate topics data with a
Hansard data file for one sitting day.

\hypertarget{code-availability}{%
\section{Code Availability}\label{code-availability}}

The code written to build this database is available on the GitHub
repository associated with this paper
(https://github.com/lindsaykatz/hansard-proj). All scripts were created
using R software\textsuperscript{21}. The core packages used to develop
these scripts are: the \texttt{XML} package\textsuperscript{15}, the
\texttt{xml2} package\textsuperscript{16}, the \texttt{tidyverse} R
packages\textsuperscript{22}, the \texttt{AustralianPoliticians}
package\textsuperscript{9}, and the \texttt{ausPH} package. \texttt{XML}
and \texttt{xml2} were used for parsing the XML documents,
\texttt{AustralianPoliticians} and \texttt{ausPH} were used for cleaning
up and filling in MP details in the datasets, and \texttt{tidyverse}
packages were used in all steps, for tidy wrangling of data.

\hypertarget{acknowledgements}{%
\section{Acknowledgements}\label{acknowledgements}}

We thank Kristine Villaluna, Monica Alexander, and Jack Stephenson for
helpful comments. We wish to especially thank Guy Jones, Chief Editor,
the two anonymous reviewers, and the Editorial Board Member, Michael
Jankowski for their comments.

\hypertarget{author-contributions}{%
\section{Author Contributions}\label{author-contributions}}

LK developed and implemented the code to obtain, create, and test, the
datasets and wrote the first draft of the paper. RA conceptualized and
designed the study, and contributed writing. Both authors approved the
final version.

\hypertarget{competing-interests}{%
\section{Competing Interests}\label{competing-interests}}

The authors declare no competing interests.

\hypertarget{references}{%
\section*{References}\label{references}}
\addcontentsline{toc}{section}{References}

\hypertarget{refs}{}
\begin{CSLReferences}{0}{0}
\leavevmode\vadjust pre{\hypertarget{ref-citehansard}{}}%
\CSLLeftMargin{1. }%
\CSLRightInline{Commonwealth of Australia.
\href{https://www.aph.gov.au/Parliamentary_Business/Hansard}{{Parliamentary
Debates, House of Representatives}}. (2023).}

\leavevmode\vadjust pre{\hypertarget{ref-vice2017history}{}}%
\CSLLeftMargin{2. }%
\CSLRightInline{Vice, J. \& Farrell, S. \emph{{The history of Hansard}}.
(House of Lords Library; House of Lords Hansard, 2017).}

\leavevmode\vadjust pre{\hypertarget{ref-lipad}{}}%
\CSLLeftMargin{3. }%
\CSLRightInline{Beelen, K. \emph{et al.} {Digitization of the Canadian
parliamentary debates}. \emph{Canadian Journal of Political
Science/Revue canadienne de science politique} \textbf{50}, 849--864
(2017).}

\leavevmode\vadjust pre{\hypertarget{ref-Erjavec2022}{}}%
\CSLLeftMargin{4. }%
\CSLRightInline{Erjavec, T. \emph{et al.}
\href{https://doi.org/10.1007/s10579-021-09574-0}{{The ParlaMint corpora
of parliamentary proceedings}}. \emph{Language Resources and Evaluation}
\textbf{57}, 415--448 (2022).}

\leavevmode\vadjust pre{\hypertarget{ref-Rauh}{}}%
\CSLLeftMargin{5. }%
\CSLRightInline{Rauh, C. \& Schwalbach, J. {The ParlSpeech V2 data set:
Full-text corpora of 6.3 million parliamentary speeches in the key
legislative chambers of nine representative democracies}. (2020)
doi:\href{https://doi.org/10.7910/DVN/L4OAKN}{10.7910/DVN/L4OAKN}.}

\leavevmode\vadjust pre{\hypertarget{ref-parlee}{}}%
\CSLLeftMargin{6. }%
\CSLRightInline{Sylvester, C., Ershova, A., Khokhlova, A., Yordanova, N.
\& Greene, Z. {ParlEE plenary speeches V2 data set: Annotated full-text
of 15.1 million sentence-level plenary speeches of six EU legislative
chambers}. \emph{Harvard Dataverse} (2023)
doi:\href{https://doi.org/10.7910/DVN/VOPK0E}{10.7910/DVN/VOPK0E}.}

\leavevmode\vadjust pre{\hypertarget{ref-maple}{}}%
\CSLLeftMargin{7. }%
\CSLRightInline{Kartalis, Y. \& Costa Lobo, M. {MAPLE Parliamentary
Datasets: Full-text and annotated corpora of parliamentary speeches in
the legislatures of six European democracies}. (2021)
doi:\href{https://doi.org/10.7910/DVN/9MN0RL}{10.7910/DVN/9MN0RL}.}

\leavevmode\vadjust pre{\hypertarget{ref-australia2018house}{}}%
\CSLLeftMargin{8. }%
\CSLRightInline{House of Representatives. \emph{{House of
Representatives Practice}} ({Australian Government - Department of the
House of Representatives}, 2018).}

\leavevmode\vadjust pre{\hypertarget{ref-auspol}{}}%
\CSLLeftMargin{9. }%
\CSLRightInline{Alexander, R. \& Hodgetts, P. A.
\emph{\href{https://CRAN.R-project.org/package=AustralianPoliticians}{{AustralianPoliticians:
Provides Datasets About Australian Politicians}}}. (2021).}

\leavevmode\vadjust pre{\hypertarget{ref-Dowding2021}{}}%
\CSLLeftMargin{10. }%
\CSLRightInline{Dowding, K., Leslie, P. \& Taflaga, M. {Australia}. in
\emph{{The Politics of Legislative Debates}} 130--151 (Oxford University
Press, 2021).
doi:\href{https://doi.org/10.1093/oso/9780198849063.003.0008}{10.1093/oso/9780198849063.003.0008}.}

\leavevmode\vadjust pre{\hypertarget{ref-salisbury2011}{}}%
\CSLLeftMargin{11. }%
\CSLRightInline{Salisbury, C. {'Mr Speaker, I withdraw...': standards of
(mis) behaviour in the Queensland, Western Australian and Commonwealth
parliaments compared via online Hansard}. \emph{Australasian
Parliamentary Review} \textbf{26}, 166--177 (2011).}

\leavevmode\vadjust pre{\hypertarget{ref-rasiah2010framework}{}}%
\CSLLeftMargin{12. }%
\CSLRightInline{Rasiah, P. {A framework for the systematic analysis of
evasion in parliamentary discourse}. \emph{Journal of Pragmatics}
\textbf{42}, 664--680 (2010).}

\leavevmode\vadjust pre{\hypertarget{ref-fraussen2018}{}}%
\CSLLeftMargin{13. }%
\CSLRightInline{Fraussen, B., Graham, T. \& Halpin, D. R. {Assessing the
prominence of interest groups in parliament: a supervised machine
learning approach}. \emph{The Journal of Legislative Studies}
\textbf{24}, 450--474 (2018).}

\leavevmode\vadjust pre{\hypertarget{ref-alexander2021}{}}%
\CSLLeftMargin{14. }%
\CSLRightInline{Alexander, R. \& Alexander, M. {The Increased Effect of
Elections and Changing Prime Ministers on Topics Discussed in the
Australian Federal Parliament between 1901 and 2018}. \emph{arXiv
preprint arXiv:2111.09299} (2021).}

\leavevmode\vadjust pre{\hypertarget{ref-XML_pkg}{}}%
\CSLLeftMargin{15. }%
\CSLRightInline{Temple Lang, D.
\emph{\href{https://CRAN.R-project.org/package=XML}{{XML: Tools for
Parsing and Generating XML Within R and S-Plus}}}. (2022).}

\leavevmode\vadjust pre{\hypertarget{ref-xml2_pkg}{}}%
\CSLLeftMargin{16. }%
\CSLRightInline{Wickham, H., Hester, J. \& Ooms, J.
\emph{\href{https://CRAN.R-project.org/package=xml2}{{xml2: Parse
XML}}}. (2021).}

\leavevmode\vadjust pre{\hypertarget{ref-Wickham2014}{}}%
\CSLLeftMargin{17. }%
\CSLRightInline{Wickham, H.
\href{https://doi.org/10.18637/jss.v059.i10}{{Tidy Data}}. \emph{Journal
of Statistical Software} \textbf{59}, (2014).}

\leavevmode\vadjust pre{\hypertarget{ref-house2021}{}}%
\CSLLeftMargin{18. }%
\CSLRightInline{House of Representatives. \emph{{A window on the house:
Practices and procedures relating to question time}}. ({Parliament of
Australia}, 2021).}

\leavevmode\vadjust pre{\hypertarget{ref-katz_2023}{}}%
\CSLLeftMargin{19. }%
\CSLRightInline{Katz, L. \& Alexander, R. {A new, comprehensive database
of all proceedings of the Australian Parliamentary Debates (1998-2022)}.
(2023)
doi:\href{https://doi.org/10.5281/zenodo.7336075}{10.5281/zenodo.7336075}.}

\leavevmode\vadjust pre{\hypertarget{ref-infosheet_parties}{}}%
\CSLLeftMargin{20. }%
\CSLRightInline{Parliament of Australia.
\href{https://www.aph.gov.au/About_Parliament/House_of_Representatives/Powers_practice_and_procedure/00_-_Infosheets/Infosheet_22_-_Political_parties}{{Infosheet
22 - Political parties}}. \emph{House of Representatives Infosheets}
(2022).}

\leavevmode\vadjust pre{\hypertarget{ref-r_software}{}}%
\CSLLeftMargin{21. }%
\CSLRightInline{R Core Team. \emph{\href{https://www.R-project.org/}{{R:
A Language and Environment for Statistical Computing}}}. (R Foundation
for Statistical Computing, 2022).}

\leavevmode\vadjust pre{\hypertarget{ref-tidyverse}{}}%
\CSLLeftMargin{22. }%
\CSLRightInline{Wickham, H. \emph{et al.}
\href{https://doi.org/10.21105/joss.01686}{{Welcome to the tidyverse}}.
\emph{Journal of Open Source Software} \textbf{4}, 1686 (2019).}

\end{CSLReferences}

\end{document}